\documentclass[conference]{IEEEtran}
\IEEEoverridecommandlockouts
\usepackage{cite}
\usepackage{amsmath,amssymb,amsfonts}
\usepackage{algorithmic}
\usepackage{graphicx}
\usepackage{textcomp}
\usepackage{xcolor}

\usepackage{academicons}
\definecolor{orcidlogocol}{HTML}{A6CE39}
\usepackage{hyperref}
\hypersetup{colorlinks}

\newcommand{\dg}{\dagger}

\usepackage{quantikz}
\usepackage{tikz-cd}
    
\begin{document}

\title{Numerical circuit synthesis and compilation for multi-state preparation\\
}

\author{\IEEEauthorblockN{Aaron Szasz}
\IEEEauthorblockA{\textit{Computational Research Division} \\
\textit{Lawrence Berkeley National Laboratory}\\
Berkeley, USA \\
aszasz@lbl.gov, \href{https://orcid.org/0000-0002-1127-2111}{\textcolor{orcidlogocol}{ID}}}
\and
\IEEEauthorblockN{Ed Younis}
\IEEEauthorblockA{\textit{Computational Research Division} \\
\textit{Lawrence Berkeley National Laboratory}\\
Berkeley, USA \\
edyounis@lbl.gov, \href{https://orcid.org/0000-0002-1306-1860}{\textcolor{orcidlogocol}{ID}}}
\and
\IEEEauthorblockN{Wibe de Jong}
\IEEEauthorblockA{\textit{Computational Research Division} \\
\textit{Lawrence Berkeley National Laboratory}\\
Berkeley, USA \\
wadejong@lbl.gov, \href{https://orcid.org/0000-0002-7114-8315}{\textcolor{orcidlogocol}{ID}}}
}

\maketitle

\begin{abstract}
Near-term quantum computers have significant error rates and short coherence times, so compilation of circuits to be as short as possible is essential.  Two types of compilation problems are typically considered: circuits to prepare a given state from a fixed input state, called ``state preparation''; and circuits to implement a given unitary operation, for example by ``unitary synthesis''.  In this paper we solve a more general problem: the transformation of a set of m states to another set of m states, which we call ``multi-state preparation''.  State preparation and unitary synthesis are special cases; for state preparation, m=1, while for unitary synthesis, m is the dimension of the full Hilbert space.  We generate and optimize circuits for multi-state preparation numerically.  In cases where a top-down approach based on matrix decompositions is also possible, our method finds circuits with substantially (up to 40\%) fewer two-qubit gates.  We discuss possible applications, including efficient preparation of macroscopic superposition (``cat'') states and synthesis of quantum channels.
\end{abstract}



\section{Introduction}

Just as classical computations are carried out by a series of gates each acting on one or two bits, digital quantum computations are described by quantum circuits in which quantum gates act on quantum bits (``qubits'')~\cite{nielsen_chuang_2010}.  Unlike in the classical case, in noisy intermediate-scale quantum (NISQ) computing~\cite{NISQ_Preskill}, significant error rates impose strong limits on the useful runtime of quantum computers.  Single qubits lose coherence over a short time, meaning that the state of a qubit becomes random and uncorrelated with neighboring qubits.  

Even more problematic are the errors due to gate operations, especially from gates acting on two (or more) qubits at a time. Typical error rates for two-qubit gates, measured by infidelity, $\epsilon$, are in the range of 0.1\% to 1\% for quantum computing architectures including trapped ions~\cite{Quantinuum_gate_fidelity, IONQ_gate_fidelity}, neutral atoms~\cite{Saffman2020, Madjarov2020, evered2023highfidelity}, silicon quantum dots~\cite{Mills2022_Si_fidelity}, and superconducting qubits~\cite{IBM_gate_fidelity, Nguyen2022, Moskalenko2022, Acharya2023short}.  Since the overall fidelity of a circuit shrinks exponentially in the number of gates $d$, scaling as $(1-\epsilon)^d$, reducing the number of two-qubit gates in a circuit is essential for making near-term quantum computing feasible and useful.

To reduce the number of gates in a circuit as much as possible, we can consider two related tasks: synthesis and compilation.  In the synthesis task, we start with a given unitary operator (or quantum state) and search for the circuit with the fewest two-qubit gates that reproduces the unitary (or state).  In the compilation task, we start with a circuit that implements the unitary or state, and we try to find a new, shorter circuit that performs the same operation as the original one.

Unitary synthesis is a widely-studied problem~\cite{Zhang2004, Shende2006, Camps2020, Davis2020QSearch, Younis2021QFASTCS, Smith2023LEAP, Cincio2021, Squander,  younis2022quantum, Ashhab2022synthesis, Sun2023}, and algorithms to carry out this task are implemented in a variety of publicly available software packages, including Qiskit~\cite{Qiskit}, Cirq~\cite{Cirq}, Tket~\cite{TKet}, and Lawrence Berkeley National Laboratory's Berkeley Quantum Synthesis Toolkit (BQSKit)~\cite{BQSKit, BQSKit_code}.  State preparation is also widely studied~\cite{Schleich1997, Bergholm2005, Plesch2011, Araujo2021, Zhang2022, Ashhab2022synthesis, Sun2023}, with algorithms again implemented in the software packages listed above.

These two synthesis/compilation tasks, state preparation and unitary synthesis, can be generalized by considering isometries.  An isometry is a norm-preserving linear map from $n_\text{in}$ to $n_\text{out}$ qubits with $n_\text{in} \leq n_\text{out}$, and it can always be implemented (in an under-determined manner for $n_\text{in}<n_\text{out}$) by an $n_\text{out}$-qubit unitary acting on $n_\text{in}$ input qubits and $(n_\text{out}-n_\text{in})$ ancillas each initialized to $|0\rangle$.  From this perspective, the state preparation problem, in which we prepare an $n$-qubit state from the all-$|0\rangle$ state, is isometry synthesis with $n_\text{in} = 0$, while unitary synthesis is the case of $n_\text{in} = n_\text{out}$.  Then intermediate problems are those with $0 < n_\text{in} < n_\text{out}$.  Prescriptive (matrix decomposition-based) numerical methods for the general isometry synthesis problem are known~\cite{knill1995approximation, Iten} and are also available in public software packages~\cite{ItenCode, Qiskit, Cirq}.

However, a further generalization is possible.  
We can view state preparation as the problem of finding a way to prepare one state from a given (simple) input state, while unitary synthesis is the problem of simultaneously preparing $2^n$ different states of $n$ qubits from the $2^n$ standard computational basis states.  From this perspective, the more general problem is the simultaneous preparation of $m$ states efficiently for some $1\leq m\leq 2^n$.  We call this problem ``multi-state preparation''. 

Multi-state preparation not only generalizes state preparation and unitary synthesis, but also encompasses isometry synthesis (with \hbox{$m=2^{n_\text{in}}$}) and controlled quantum state preparation (CQSP)~\cite{Yuan2023optimalcontrolled}.  Furthermore, as we discuss in Sec.~\ref{sec:applications} below, multi-state preparation allows for reductions in circuit depth for a variety of practical applications, in some cases beyond what is possible via any of these four more specialized methods.

In this paper, we first define the multi-state preparation problem formally in Section~\ref{sec:multi-state_def}.  In Section~\ref{sec:sol_exist} we provide (with rigorous proof) the conditions to determine whether a multi-state preparation problem has a solution.  In Section~\ref{sec:cost_fct} we show how the problem can be solved numerically in practice; in subsection \textit{A.} we review our general approach to numerical optimization of circuits, namely circuit templating and instantiation, then in \textit{B.} we present a good cost function for the multi-state preparation problem in particular.  In \textit{C.}, we provide a detailed mathematical motivation for the cost function; readers primarily interested in using multi-state preparation as a practical tool could skip this section on the first read.  In Section~\ref{sec:applications}, we discuss several applications of multi-state preparation.  In Section~\ref{sec:examples} we demonstrate the implementation of the numerical optimization procedure in BQSKit.  We provide code for a simple example, and we show numerical results on random 3- and 4-qubit mappings, where we outperform other available software packages.  We conclude in Section~\ref{sec:discussion}.


\section{The multi-state preparation problem}\label{sec:multi-state_def} 

We begin with a formal definition.

\begin{quote}
\noindent \textit{Definition of multi-state preparation:} Given two sets of $m$ pure states of $n \geq \log_2(m)$ qubits, $\{|v_1\rangle,\cdots,|v_m\rangle\}$ and $\{|w_1\rangle,\cdots,|w_m\rangle\}$, find a quantum circuit that maps $|v_i\rangle$ to $|w_i\rangle$ for all $i$.  (The action of the circuit on states not in the span of $\{|v_1\rangle,\cdots,|v_m\rangle\}$ is not constrained.)
\end{quote}

\noindent The multi-state preparation problem includes both state preparation and unitary synthesis as special cases.  

\textit{State preparation} is the problem of finding a circuit that prepares an $n$-qubit pure state, $|\psi\rangle$, from the input product state $|0\rangle^{\otimes n}$.  Thus state preparation is a special case of multi-state preparation with $m=1$, $|v_1\rangle = |0\rangle^{\otimes n}$, and $|w_1\rangle=|\psi\rangle$.  

\textit{Unitary synthesis} is the problem of finding a circuit that implements an $n$-qubit unitary operator, $U$.  However, any unitary is precisely specified by its actions on the $2^n$ computational basis states, $|0\rangle^{\otimes n}$, $|1\rangle\otimes|0\rangle^{\otimes n-1}$, $\cdots$, $|1\rangle^{\otimes n}$; when viewed as a matrix, the images of these states written in the computation basis are precisely the columns of $U$.  Thus letting $\{|v_i\rangle\}$ be the computation basis states and $\{|w_i\rangle\}$ their images, $|w_i\rangle = U|v_i\rangle$, we see that unitary synthesis is multi-state preparation with $m=2^n$.

Synthesis of isometries is also a special case of multi-state preparation.  An isometry is defined by the mapping of the computational basis states on $n_\text{in}$ qubits to some set of $m=2^{n_\text{in}}$ states on $n \geq n_\text{in}$ qubits.  We can then append $n-n_\text{in}$ ancillas, each in the state $|0\rangle$, to each of the input states. The result is a multi-state preparation problem with $m$ input states and $m$ output states, each of $n$ qubits.

One important note is that in each of these three cases, state preparation, unitary synthesis, and isometry synthesis, the input states are all orthogonal to one another.  As we will see in Sec.~\ref{sec:sol_exist} below, orthogonality guarantees the existence of a solution to a multi-state preparation problem.  However, it is not a necessary condition.  Thus our statement of the multi-state preparation problem does not make any assumption on the inner products $\langle v_i|v_j\rangle$.

\subsection{Two-qubit examples}

\noindent \textit{Example 1}: Let
\begin{align}
    &|v_1\rangle = |00\rangle, \,\,\,\, |v_2\rangle = |01\rangle \label{eq:2-qubit_ex_0}\\
    &|w_1\rangle = |00\rangle, \,\,\, |w_2\rangle = |01\rangle.\nonumber
\end{align}
This apparently trivial example actually illustrates precisely why a dedicated approach to multi-state preparation is useful.  In the standard basis, this example problem amounts to specifying the first two columns of the $4\times4$ unitary matrix that a quantum circuit should implement.  We could imagine solving this problem by first appending two additional orthonormal columns, then performing a full unitary synthesis.  

If we were lucky and picked $|10\rangle\mapsto|10\rangle$ and $|11\rangle\mapsto|11\rangle$, giving the identity transformation as the full unitary, we would find a circuit with no gates whatsoever, evidently the shortest possible circuit solving the original problem.  On the other hand, if we happened to fill out the rest of the unitary transformation as $|10\rangle\mapsto|11\rangle$ and $|11\rangle\mapsto|10\rangle$, the full unitary would be given by a single CNOT with the first qubit as the control.

In other words, each multi-state preparation problem (with $m<2^n$) corresponds to many unitary synthesis problems, and some will have shorter circuits as solutions than others.  From this perspective, in solving a multi-state preparation problem, we aim to find the optimal unitary transformation for the given state mapping as well as the shortest implementation of that unitary.  Using the numerical approach described in Section~\ref{sec:cost_fct}, we perform both optimizations simultaneously. 

\vspace{0.3cm}

\noindent \textit{Example 2}: Let 
\begin{align}
    &|v_1\rangle = |00\rangle, \,\,\,\, |v_2\rangle = |11\rangle\label{eq:2-qubit_ex_1}\\
    &|w_1\rangle = |00\rangle, \,\,\, |w_2\rangle = \frac{|01\rangle - |10\rangle}{\sqrt{2}}.\nonumber
\end{align}
An optimal circuit (in terms of two-qubit gate count) for this example looks like 
\begin{align}
    &\begin{tikzcd}[ampersand replacement=\&]
    \& \ctrl{1} \& \targ{} \& \qw \\
    \& \gate{H} \& \ctrl{-1} \& \qw 
    \end{tikzcd} \label{eq:ex1_circuit_CH}\\
    = &
    \begin{tikzcd}[ampersand replacement=\&]
    \& \qw \& \ctrl{1} \& \gate{R_y\left(\frac{3\pi}{2}\right)} \& \ctrl{1} \& \gate{R_y\left(\frac{3\pi}{2}\right)} \& \qw \\
    \& \gate{R_y\left(\frac{7\pi}{4}\right)} \& \targ{} \& \gate{R_y\left(\frac{7\pi}{4}\right)} \& \targ{} \& \gate{R_y\left(\frac{\pi}{2}\right)} \& \qw
    \end{tikzcd}\label{eq:ex1_circuit_CNOT}
\end{align}

Notably, two 2-qubit gates are required.  In comparison, one could imagine solving this problem using a unitary synthesis method, which would require first adding two more orthonormal output states for inputs $|01\rangle$ and $|10\rangle$; however, generic two-qubit unitaries require three CNOTs~\cite{Vidal2004, Vatan2004}, so in general we get a shorter circuit by doing the multi-state preparation problem directly.  On the other side, single-state preparation for two qubits uses at most one CNOT gate~\cite{Znidaric2008}, so all but a set of measure 0 of multi-state preparation problems cannot be solved by single-state preparation. Thus there are situations in which multi-state preparation will be beneficial relative to both of the more common synthesis/compilation problems.

\vspace{0.3cm}

\noindent \textit{Example 3}: As noted above, the states can also be non-orthogonal.  For example, with $|v_i\rangle$, $|w_i\rangle$ as in Eq.~\ref{eq:2-qubit_ex_1}, let
\begin{align}
    &|\tilde{v}_2\rangle = \frac{|v_1\rangle + |v_2\rangle}{\sqrt 2}\label{eq:2-qubit_ex_2}\\
    &|\tilde{w}_2\rangle = \frac{|w_1\rangle + |w_2\rangle}{\sqrt 2}.\nonumber
\end{align}
Then the exact same circuits from \eqref{eq:ex1_circuit_CH} and \eqref{eq:ex1_circuit_CNOT} will also solve the problem $|v_1\rangle \mapsto |w_1\rangle$, $|\tilde{v}_2\rangle \mapsto |\tilde{w}_2\rangle$.

\vspace{0.3cm}

\noindent \textit{Example 4}: As discussed in the next section, some problems with non-orthogonal states have no solution.  For example, let
\begin{align}
    &|v_1\rangle = |00\rangle, \,\,\,\, |v_2\rangle = \frac{|00\rangle+|11\rangle}{\sqrt 2}\label{eq:2-qubit_ex_3}\\
    &|w_1\rangle = \frac{|00\rangle + |01\rangle}{\sqrt 2}, \,\,\, |w_2\rangle = \frac{|01\rangle + |10\rangle}{\sqrt{2}}.\nonumber
\end{align}
Note that $\langle v_1|v_2\rangle=1/\sqrt{2}$ while $\langle w_1|w_2\rangle=1/2$.  As we show below, this implies that there is no solution to this particular multi-state mapping problem.


\section{When does a multi-state preparation problem have a solution in principle?}\label{sec:sol_exist}

Given two sets of $m$ n-qubit states each, $\{|v_i\rangle\}$ and $\{|w_i\rangle\}$, we want to find a quantum circuit that maps $|v_i\rangle$ to $|w_i\rangle$ for all $i\in\{1,\cdots,m\}$.  The first question to ask is whether a solution exists at all.  Fortunately, there is a simple answer.\footnote{Note that this result can also be found on the Mathematics Stack Exchange forum~\cite{MathStackExch_existence}.}

\begin{quote}
    \noindent \textit{Solution}: Let $V$ be the matrix whose columns contain the states $|v_i\rangle$, represented as vectors in the computational basis, and likewise let $W$ be a matrix with columns $|w_i\rangle$.  Both matrices are $2^n \times m$. A circuit that maps $|v_i\rangle$ to $|w_i\rangle$ $\forall i$ exists if and only if $V^\dagger V = W^\dagger W$, i.e. if and only if $\langle v_i|v_j\rangle = \langle w_i|w_j\rangle$ for all $i$, $j$.
\end{quote}

\begin{figure}
    \centering
    \includegraphics[width=\columnwidth]{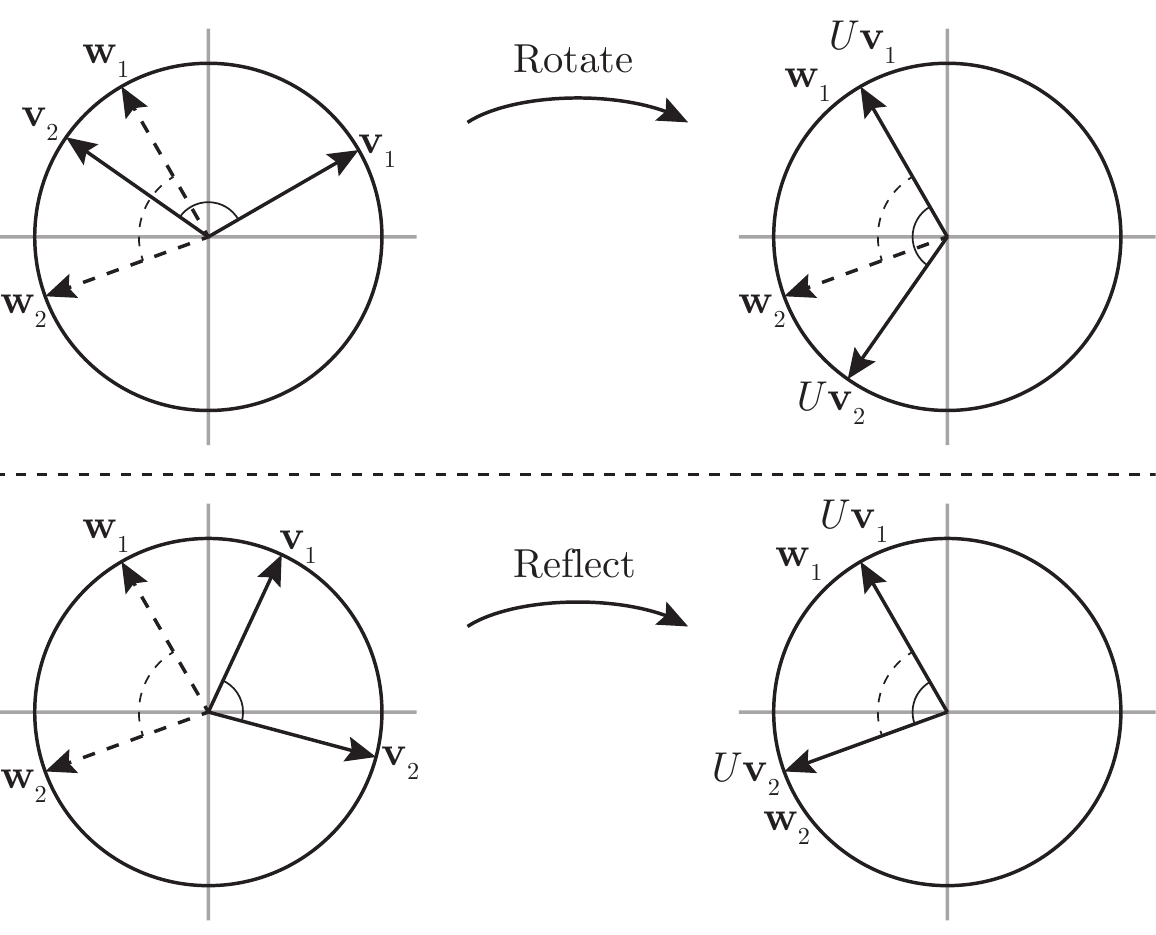}
    \caption{There exists a unitary that maps vectors $(\mathbf{v}_1,\mathbf{v}_2)$ to $(\mathbf{w}_1,\mathbf{w}_2)$ if and only if the angles between the vectors are the same, $\mathbf{v}_1\cdot \mathbf{v}_2 = \mathbf{w}_1\cdot\mathbf{w}_2$.  In the top row, we see that when the angles are different, a unitary $U$ that takes $\mathbf{v}_1$ to $\mathbf{w}_1$ cannot also take $\mathbf{v}_2$ to $\mathbf{w}_2$.  Conversely, on the bottom is an example where the angles match, and thus there exists a unitary that maps both vectors simultaneously.  This is the intuition behind the condition $V^\dagger V = W^\dagger W$ for solvability of a multi-state preparation problem.}
    \label{fig:circle_analogy}
\end{figure}

\noindent \textbf{Intuition}: By analogy, we consider four vectors on the unit circle, $\mathbf{v}_1$, $\mathbf{v}_2$, $\mathbf{w}_1$, and $\mathbf{w}_2$, and we ask whether there is some rotation plus reflection that takes $\mathbf{v}_1$ to $\mathbf{w}_1$ and $\mathbf{v}_2$ to $\mathbf{w}_2$.  The situation is illustrated in Fig.~\ref{fig:circle_analogy}.

We can of course always find a rotation that maps $\mathbf{v}_1$ to $\mathbf{w}_1$.  Once we carry out this rotation, it becomes visually clear whether or not we can simultaneously map $\mathbf{v}_2$ to $\mathbf{w}_2$: as shown in the figure, this is possible only when the angle between $\mathbf{v}_1$ and $\mathbf{v}_2$ is the same (up to sign) as the angle between $\mathbf{w}_1$ to $\mathbf{w}_2$.  Intuitively, this is because rotations and reflections preserve angles, so one set of vectors can be transformed to another if and only if the angles between the initial and between the final vectors match.

Returning to our multi-state preparation problem, the rotations and reflections become the unitary action of the desired quantum circuit, while the angles between vectors correspond to the overlaps $\langle v_i|v_j\rangle$ and $\langle w_i|w_j\rangle$.  These are precisely contained in the matrices $V^\dagger V$ and $W^\dagger W$:
\begin{equation}
    V^\dagger V = \left(\begin{array}{cccc}
    \langle v_1|v_1\rangle & \langle v_1|v_2\rangle & \cdots & \langle v_1|v_m\rangle\\
    \langle v_2|v_1\rangle & \langle v_2|v_2\rangle & \cdots & \langle v_2|v_m\rangle\\
    \vdots & \vdots & \ddots & \vdots\\
    \langle v_m|v_1\rangle & \langle v_m|v_2\rangle & \cdots & \langle v_m|v_m\rangle
    \end{array}\right).
\end{equation}

\noindent \textbf{Proof}: We now prove the result.

\textit{If a circuit exists, $V^\dg V=W^\dg W$}: We first suppose that a circuit for the multi-state preparation indeed exists.  This circuit implements a unitary operator $U$, which by assumption realizes the mapping $|w_i\rangle = U |v_i\rangle$.  Then 
\begin{equation}
    \langle w_i|w_j\rangle = \langle U v_i|U v_j\rangle = \langle v_i|U^\dg U v_j\rangle = \langle v_i|v_j\rangle
\end{equation}
and hence each element of the two matrices is equal and $W^\dg W = V^\dg V$. \hfill $\blacksquare$

\textit{If $V^\dg V=W^\dg W$, a circuit exists}: We now suppose that $V^\dagger V = W^\dagger W$.  Let's call this matrix $O$, for ``overlap''.  $O$ is Hermitian by construction, hence diagonalizable.  Furthermore, $O$ is positive semi-definite: $\langle \psi| V^\dagger V |\psi\rangle = \langle V\psi | V\psi \rangle = |\!|V\psi|\!|^2 \geq 0$.  We thus let $O=S D S^\dg$ where $S$ is unitary and $D$ is diagonal with only positive real entries.  ($D$ is only guaranteed to be nonnegative, but if $D$ has zero entries, we can just remove them and remove the corresponding columns of $S$ to get smaller matrices, making $D$ strictly positive.  Note that $D$ will have zero entries when the columns of $V$ are linearly dependent.)

Then let $\tilde{V} = V S D^{-1/2}, \tilde{W} = W S D^{-1/2}$.  Intuitively, $VS$ is the matrix whose columns are derived by orthogonalizing the columns of $V$, and the factor of $D^{-1/2}$ normalizes the vectors to be orthonormal.  Then both $\tilde{V}$ and $\tilde{W}$ are isometries:
\begin{align}
    \tilde{V}^\dagger\tilde{V} &= D^{-1/2} S^\dagger V^\dagger V S D^{-1/2} = D^{-1/2} S^\dagger O S D^{-1/2} \nonumber
    \\&= D^{-1/2} D D^{-1/2} = \text{Id}
\end{align}
and likewise for $\tilde{W}$.  In general, $\tilde{V}$ and $\tilde{W}$ have more rows than columns: they are isometries from a smaller space into a larger one.  In that case, we append further orthogonal columns to make them into square matrices while maintaining the isometric conditions $\tilde{V}^\dg \tilde{V}=\text{Id}$ and $\tilde{W}^\dg \tilde{W}=\text{Id}$.  Any square isometry is unitary, so both $\tilde{V}$ and $\tilde{W}$ are then unitary.

Finally, let $U=\tilde{W}\tilde{V}^\dagger$.  As a product of two unitaries, $U$ is also unitary.

Furthermore, if we didn't need to add extra columns to $\tilde{V}$ and $\tilde{W}$,
\begin{align}
    UV &= WSD^{-1}S^\dagger V^\dagger V = WSD^{-1}S^\dagger O\nonumber \\&= WSD^{-1}S^\dagger S D S^\dagger = W.
\end{align}
If we did add extra columns, then letting the matrices containing just the extra columns be $C_V$ and $C_W$ respectively, we have
\begin{align}
    UV = W + C_W C_V^\dg V,
\end{align}
but all the columns of $V$ are in the span of the columns of $VSD^{-1/2}$ and the columns in $C_V$ were chosen to be orthogonal, so $C_V^\dg V = 0$.

Thus given $V^\dagger V = W^\dagger W$, we have explicitly constructed a unitary that maps $V$ to $W$ and hence $|v_i\rangle$ to $|w_i\rangle$ for all $i$. 

\hfill $\blacksquare$


\section{Numerical implementation}\label{sec:cost_fct}

Now that we know how to tell whether a solution to a multi-state preparation problem exists, we turn to the question of finding that solution in practice.  We consider the framework of synthesis and compilation using instantiation of parameterized circuit templates with numerical optimization~\cite{younis2022quantum}.
In this framework, we find an optimal circuit using a cost function that is minimized when the circuit solves the problem of interest.

Here we first briefly review parameterized circuit instantiation, then we present and motivate a cost function for the multi-state preparation problem.

\subsection{Background: numerical optimization via instantiation}

Suppose we want to find a quantum circuit that solves some task, for example preparing a given state from $|0\rangle^{\otimes n}$ or synthesizing a unitary.  We perform synthesis by repeating two steps:
\begin{enumerate}
    \item Generate a circuit ``template'' containing a variety of parameterized gates such as $\exp(i\theta Z)$, with unspecified parameters.
    \item Given this circuit template, find the parameter values that bring the circuit as close as possible to a solution to the problem of interest.  If the best possible parameterization fails to produce a circuit that solves the problem, go back to step 1.
\end{enumerate}
Step 2 requires evaluating, for each set of parameters, how close the parameterized circuit comes to solving the problem (e.g. preparing the desired state).  We perform this evaluation by means of a problem-specific cost function.  

For example, for unitary synthesis, if $U_C$ is the unitary operator implemented by the circuit and $U$ is the desired unitary, a good cost function is $1-|\text{Tr}[U_C U^\dagger]|/2^n$~\cite{younis2022quantum}, which takes values in $[0,1]$ and is minimized at 0 if and only if $U_C=U$.  For preparation of a state $|\psi\rangle$, a good cost function is $1-|\langle\psi|U_C|\mathbf{0}\rangle|$ or equivalently $1-|\text{Tr}[U_C |\mathbf{0}\rangle\langle\psi|]|$,\footnote{We use $|\mathbf{0}\rangle$ as shorthand for $|0\rangle^{\otimes n}$} which is again minimized at 0 for a correct solution.

For a given cost function, there remain many possible approaches to generating circuit templates and to carrying out the parameter optimization.  BQSKit includes a variety of strategies, including QSearch~\cite{Davis2020QSearch}, QFAST~\cite{Younis2021QFASTCS}, and LEAP~\cite{Smith2023LEAP}.  The circuit templates can also be generated in a way that respects the connectivity of some particular quantum hardware~\cite{Weiden2022}.

\subsection{Cost function for multi-state preparation}
Consider the matrices $V$ and $W$ introduced in Sec.~\ref{sec:sol_exist}, each of whose $m$ columns are $n$-qubit states.  Then a natural choice of cost function for multi-state preparation is:
\begin{equation}
    1-\frac{1}{m}|\text{Tr}[U_C VW^\dagger]|.\label{eq:correct_cost_fct}
\end{equation}
This cost function includes the ones for state preparation and for unitary synthesis as special cases:  
\begin{itemize}
    \item For state preparation, $V$ is the column vector $(1,0,\cdots,0)^\text{T}$, while $W$ is the column vector containing the elements of $|\psi\rangle$ in the computational basis.  Then the cyclic property of the trace gives $\text{Tr}[U_C VW^\dagger] = \langle \psi|U_C|\mathbf{0}\rangle$.  

    \item For unitary synthesis, $V$ is just the identity matrix, so that its columns are the successive computational basis vectors, and $W=U$ since its columns are the states mapped to by said basis vectors.  Also noting that $m=2^n$, we exactly recover the cost function for synthesis.
\end{itemize}

For any multi-state preparation problem, the cost function~\eqref{eq:correct_cost_fct} lies in the range $[0,1]$ and has a minimum value at 0 for a correct solution: in that case, $U_C V = W$, and 
\begin{equation}
    \text{Tr}[WW^\dg] = \text{Tr}[W^\dg W] = \sum_i \langle w_i|w_i\rangle = m.
\end{equation}
Note that this proof does not require orthogonality of the $|w_i\rangle$, or even linear independence.

\subsection{Motivation for cost function}

Suppose that the state system in question has a solution, $U_C$, which precisely maps $|v_i\rangle$ to $|w_i\rangle$.  Then for all $i$, $\langle w_i |U_C |v_i\rangle=1$.  For arbitrary $U$, we have $|\langle w_i|U|v_i\rangle| \leq 1$, so any cost function of the form 
\begin{equation}
1-\left|\sum_i a_i\langle w_i|U|v_i\rangle\right| \label{eq:cost_fct_weights_version}
\end{equation}
for $a_i > 0$, $\sum_i a_i=1$ will satisfy 
\begin{equation}
1-\left|\sum_i a_i\langle w_i|U|v_i\rangle\right| \geq 1-\sum_i a_i|\langle w_i|U|v_i\rangle| \geq 1 -\sum_i a_i = 0
\end{equation}
and will achieve the minimum value of 0 precisely for correct solutions $U_C$.  

The cost function given in Eq.~\eqref{eq:correct_cost_fct} is of the form~\eqref{eq:cost_fct_weights_version}:
\begin{equation}
    \frac{1}{m}\text{Tr}[U VW^\dagger] = \frac{1}{m}\sum_{i} \text{Tr}[U |v_i\rangle\langle w_i|] = \sum_i \frac{1}{m} \langle w_i | U |v_i\rangle. 
\end{equation}
However, the choice of $a_i=1/m$ is not obvious and deserves some justification.  

First, suppose that all the input (equivalently, output) vectors form an orthonormal set, $V^\dagger V = \text{Id}$.  Then all pairs of vectors $(|v_i\rangle, |w_i\rangle)$ are equivalent under some high-dimensional reflection and rotation, so it is natural to weight the mapping of each pair equally.  $1/m$ is thus a natural weight to assign.

What about the more general case where the provided input and output vectors are not orthogonal?  Consider the following example, which we will return to several times:

\vspace{0.1cm}

\begin{quote}
\noindent \textit{Example}:\label{ex:non-ortho_two_states} Let $|\psi_1\rangle$ and $|\psi_2\rangle$ be orthogonal states, and define $|v_\pm\rangle = \sqrt{1-\epsilon^2}|\psi_1\rangle \pm \epsilon|\psi_2\rangle$ for some small $\epsilon>0$.  So the two specified input states (and, if the problem is solvable, the corresponding output states) are nearly parallel, with 
\begin{equation}
    V^\dagger V = \left(\begin{array}{cc}
    1 & 1-2\epsilon^2 \\
    1-2\epsilon^2 & 1 
    \end{array}\right).
\end{equation}
The desired images of the input states are $|v_\pm\rangle\mapsto |w_\pm\rangle$ and $|\psi_i\rangle\mapsto|\phi_i\rangle$.
\end{quote}
We will keep this example in mind as we proceed to derive two possible cost functions and then compare them.

An exact solution to the multi-state preparation problem would have each state in the span of $\{|v_i\rangle\}$ mapped to the same linear combination of the $\{|w_i\rangle\}$, so a natural way to define the total cost function for the system is to average over the error in mapping each individual state in Span($\{|v_i\rangle\}$).  Consider some such state $|v\rangle$, which in the computational basis we write as $V\mathbf{c}$, for some coefficient vector $\mathbf{c}$; this state should map to $W\mathbf{c}$, so for a given circuit implementing the unitary $U_C$, the error for this state is $|\!|V\mathbf{c}|\!|^2-\mathbf{c}^\dagger W^\dagger U_C V\mathbf{c}$.  (If the columns of $V$ are orthonormal, $|\!|V\mathbf{c}|\!|^2=1$, but that is not true in general.) Thus we could write the overall cost for the circuit as an average over different coefficient vectors $\mathbf{c}$:
\begin{equation}
    \Delta = \frac{1}{\mathcal{V}}\int_\mathbf{c}\!\mathbf{c}^\dg O \mathbf{c}\,d\mathbf{c} - \frac{1}{\mathcal{V}}\int_\mathbf{c} \mathbf{c}^\dagger W^\dagger U_C V\mathbf{c}\,\, d\mathbf{c}\label{eq:state_avg_generic}
\end{equation}
where $\mathcal{V}$ measures the volume of Span($\{|v_i\rangle\}$) and $O$ is the overlap matrix $V^\dg V$.  Of course, this is not yet well-defined, since we need an integration measure on the coefficient vectors.  This choice of integration measure effectively assigns different weights to the states in Span($\{|v_i\rangle\}$) and is what determines the weights $a_i$ in Eq.~\eqref{eq:cost_fct_weights_version}.  We take two different approaches.

\textit{Approach 1}: First, we could sample $\mathbf{c}$ as random $m$-dimensional vectors, i.e. from the first column of $m\times m$ Haar-random unitaries; for unitary $U_m$, the first column is $U_m\mathbf{e}_1$ with $\mathbf{e}_1=(1,0,\cdots,0)^\text{T}$.  Then Eq.~\eqref{eq:state_avg_generic} becomes 
\begin{equation}
    \mathbf{e}_1^T \left[\int_\text{Haar}\!\!\!\! U_m^\dg O U_m\,d\mu(U_m) - \int_\text{Haar}\!\!\!\! U_m^\dagger W^\dagger U_C V U_m\, d\mu(U_m)\right] \mathbf{e}_1
\end{equation}
where $d\mu$ is the Haar measure.  The Haar average can be computed exactly, by the ``twirling'' integral~\cite{Roberts2017}
\begin{equation}
    \int_\text{Haar} U^\dagger A U \,d\mu(U) = \frac{\text{Tr}[A]}{m}\mathbf{Id}.
\end{equation}
Using $\text{Tr}[O]=m$, we thus get a cost function of
\begin{equation}
    \left[1-\frac{1}{m}\text{Tr}[W^\dagger U_C V] \right]\,\left(\mathbf{e}_1^T \cdot \text{Id}\cdot\mathbf{e}_1\right) = 1-\frac{1}{m}\text{Tr}[W^\dagger U_C V],
\end{equation}
which is precisely Eq.~\eqref{eq:correct_cost_fct} up to taking the magnitude, which we do in \eqref{eq:correct_cost_fct} only for the convenience of working with real numbers in numerical optimization.

We claimed that our choice of distribution for sampling the coefficients $\mathbf{c}$ corresponds to a choice of weights for the average over states in Span($\{|v_i\rangle\}$).  To understand the weights that arise from unitary sampling of $\mathbf{c}$ as considered here, we turn back to the example above.  The cost function becomes
\begin{equation}
    1-\left(\langle w_+|U_C|v_+\rangle + \langle w_-|U_C|v_-\rangle\right)/2
\end{equation}
which in terms of the orthogonal states $|\psi_1\rangle$ and $|\psi_2\rangle$ is 
\begin{equation}
    1 - \left((1-\epsilon^2)\langle \phi_1|U_C|\psi_1\rangle + \epsilon^2\langle \phi_2|U_C|\psi_2\rangle\right).
\end{equation}
So in the cost function, the orthogonal basis states for Span($\{|v_i\rangle\}$) are effectively weighted by their prevalence in the specified input states.  In this case, most of the weight is given to $|\psi_1\rangle$ since that is also the case for $|v_\pm\rangle$.  

To get the corresponding general result, we recall that the overlap matrices can be diagonalized as $V^\dagger V = W^\dagger W = S D S^\dagger$.  Then the cost function can be rewritten with
\begin{equation}
    \text{Tr}[W^\dagger U V] = \text{Tr}[D \left(W S D^{-1/2}\right)^\dg U \left(VSD^{-1/2}\right)].
\end{equation}
Letting $|\tilde{v}_i\rangle$ and $|\tilde{w}_i\rangle$ be the (orthonormal) columns of $VSD^{-1/2}$ and $WSD^{-1/2}$, we can then rewrite the cost function~\eqref{eq:correct_cost_fct} as 
\begin{equation}
    1-\frac{1}{m}\left|\sum_i d_i \langle \tilde{w}_i|U_C|\tilde{v}_i\rangle\right|.
\end{equation}
Thus in terms of this orthonormal basis for Span($\{|v_i\rangle\}$), the weights are given by the eigenvalues of $V^\dg V$, which are the weights of the $|\tilde{v}\rangle$ basis states in the original input vectors:
\begin{equation}
    d_i = \sum_j |\langle v_j|\tilde{v}_i\rangle|^2.
\end{equation}

\vspace{0.1cm}

\textit{Approach 2}: Alternatively, we could sample $V\mathbf{c}$ as random $m$-dimensional vectors from the first column of $m\times m$ Haar-random unitaries.  Conceptually, this corresponds to first finding Span($\{|v_i\rangle\}$), then forgetting about the original input states and just averaging over the space in a uniform way; in practice, we can take linear combinations given by $U_m\mathbf{e}_1$ over some orthonormal basis of the span, for example the columns of $VSD^{-1/2}$.  Then Eq.~\eqref{eq:state_avg_generic} becomes 
\begin{equation}
    1-\mathbf{e}_1^T \left[\int_\text{Haar}\!\!\!\! U_m^\dagger (WSD^{-1/2})^\dagger U_C (VSD^{-1/2}) U_m\, d\mu(U_m)\right] \mathbf{e}_1.
\end{equation}
This gives
\begin{equation}
    1-\frac{1}{m}\text{Tr}[(WSD^{-1/2})^\dagger U_C (VSD^{-1/2})].\label{eq:cost_fct_alternate}
\end{equation}
In the style of Eq.~\eqref{eq:cost_fct_weights_version}, we have
\begin{equation}
    \Delta = 1-\frac{1}{m}\left|\sum_i \langle \tilde{w}_i|U_C|\tilde{v}_i\rangle\right|.
\end{equation}

\vspace{0.1cm}

\textit{Which approach is best?} Both approaches seem reasonable, a priori, but Approach 1 is preferred for several reasons:
\begin{itemize}
    \item Approach 1 is more flexible as a software tool.  If we implemented Approach 2, an end-user would have no way to access Approach 1, since the code would automatically perform the orthogonalization and re-weighting, so that the user could not choose to specify other weights if desired.  Conversely, a user could separately perform diagonalization of $V^\dagger V$ and input $VSD^{-1/2}$ and $WSD^{-1/2}$ in place of $V$ and $W$, thus recreating Approach 2 from Approach 1.

    \item Computing Eq.~\eqref{eq:cost_fct_alternate} requires inverting the overlap matrix, so if the input states are nearly linearly dependent as in the example above with a small $\epsilon$, the cost function involves a numerically unstable operation.
    
    \item One might want to find a circuit that gives a good approximate mapping in the case that $V^\dagger V \neq W^\dagger W$.  The cost function of Eq.~\eqref{eq:correct_cost_fct} remains well-defined and corresponds to an intuitive picture of how an approximate mapping should behave.

    \setlength{\parindent}{1em}
    Consider again our example problem, but now suppose that the requested mapping is 
    \begin{align}
        |v_\pm\rangle &= \sqrt{1-\epsilon^2}|\psi_1\rangle \pm \epsilon|\psi_2\rangle \\
        \mapsto |w_\pm\rangle &= \sqrt{1-4\epsilon^2}|\psi_1\rangle \pm 2\epsilon |\psi_2\rangle.
    \end{align}
    Evidently, there is no exact solution, but for small $\epsilon$ the identity transformation is close.  Specifically, using Eq.~\eqref{eq:correct_cost_fct}, we find an appropriately small cost of $\approx \epsilon^2/2$.  
    
    On the other hand, Eq.~\eqref{eq:cost_fct_alternate} is no longer well-defined.  We could define $VSD^{-1/2}$ using $S$ and $D$ from $V^\dagger V$ and $WSD^{-1/2}$ using $S$ and $D$ from $W^\dg W$, but then in the example we would find a cost of exactly 0, a clearly nonsensical result.
    \setlength{\parindent}{0em}
\end{itemize}


\section{Applications}\label{sec:applications}

Multi-state preparation has a variety of practical applications that go beyond single-state preparation and unitary synthesis.  We address three such applications in some detail below.  Other applications (for which we would use multi-state preparation as a subroutine in CQSP~\cite{Yuan2023optimalcontrolled}) include solving linear systems~\cite{Wossnig2018} and quantum clustering algorithms~\cite{Kerenidis2019}.

\subsection{Preparation of macroscopic superposition states}

Multi-state preparation on a few qubits is helpful as a subroutine for preparing a single highly-entangled state of many qubits.  Imagine we want to prepare a macroscopic superposition state (often called a ``cat'' state) of the form
\begin{equation}
    |\psi\rangle = \frac{1}{\sqrt 2} \left(|\phi_1\rangle + |\phi_2\rangle\right)
\end{equation}
where $|\phi_1\rangle$ and $|\phi_2\rangle$ are orthogonal $n$-qubit states.  

An important special case is where the system can be divided into few-qubit pieces, and between each piece there is no entanglement in either $|\phi_1\rangle$ or $|\phi_2\rangle$, and furthermore where for each few-qubit piece, the local states of $|\phi_1\rangle$ and $|\phi_2\rangle$ are orthogonal.  This class of states sounds contrived, but it includes some very important examples both for theory and for practical simulation of physical systems:
\begin{itemize}
    \item The Greenberger-Horne-Zeilinger (GHZ) state: $|\psi\rangle = (|0\rangle^{\otimes n} + |1\rangle^{\otimes n})/\sqrt{2}$.  This is the simplest example, where each of the two pieces is a product over single-qubit states. 

    \item Any state of the form
        \begin{equation}
            |\psi\rangle = \frac{1}{\sqrt{2}}\left((a|00\rangle + b|11\rangle)^{\otimes n/2} + (c|01\rangle+d|10\rangle)^{\otimes n/2}\right)\label{eq:two-qubit_chunk_example}
        \end{equation}
        Here the system is divided into two-qubit pieces, and on each piece $|\phi_1\rangle$ has a factor $(a|00\rangle + b|11\rangle)$ while $|\phi_2\rangle$ has a factor $(c|01\rangle + d|10\rangle)$.  

    \item More general states similar to~\eqref{eq:two-qubit_chunk_example}.  For example, for the first two qubits, $|\phi_1\rangle$ could have a factor $|01\rangle$ and $|\phi_2\rangle$ a factor $|10\rangle$, while for the next three, $|\phi_1\rangle$ could have a factor $(|000\rangle+|111\rangle)/\sqrt{2}$ and $|\phi_2\rangle$ a factor $(|001\rangle + |010\rangle+|100\rangle)/\sqrt{3}$.  The key requirement is that the state is divided into chunks of qubits, where within each of $|\phi_1\rangle$ and $|\phi_2\rangle$ there is no entanglement between chunks, and where for each chunk the states on $|\phi_1\rangle$ and $|\phi_2\rangle$ are orthogonal.
\end{itemize}
The GHZ state can be prepared in a constant-depth circuit (not scaling with system size) by using measurement and reset in addition to unitary operations~\cite{Briegel2001}.  We can then combine the GHZ state with multi-state preparation to immediately prepare a state like~\eqref{eq:two-qubit_chunk_example}.  We solve the multi-state preparation problem for the system
\begin{align}
    |00\rangle &\mapsto a|00\rangle + b|11\rangle\\
    |11\rangle &\mapsto c|01\rangle + d|10\rangle
\end{align}
and then just append that circuit to the one for GHZ preparation on each of the $n/2$ two-qubit blocks.

Precisely this kind of superposition is useful for measuring expectation values of operators on a quantum computer, with applications in physics and chemistry.  Suppose, for example, that one has an initial state $|\psi_0\rangle$ and wants to find the expectation value of the time-evolution operator for a Hamiltonian $H$, $U=e^{-iHt}$.  This expectation value is important in algorithms such as variational quantum phase estimation (VQPE) for finding ground states~\cite{Klymko2022_VQPE}.  A common approach to finding both the magnitude and phase of $\langle \psi_0|U|\psi_0\rangle$ is the Hadamard test, but since all operations in the state preparation and application of $U$ become controlled operations with an ancilla, the circuit depth becomes much longer than for running $U$ on its own, a major problem for near-term devices.  

A different approach, proposed in \cite{Cortes2022}, makes use of a reference state $|\text{Ref}\rangle$, one which is easy to prepare, is orthogonal to $|\psi_0\rangle$, and is an eigenstate of $H$ with known eigenvalue $E$.  In that case, the expectation value $\langle \psi_0|U|\psi_0\rangle=re^{i\theta}$, including phase, can be computed using $|\langle \psi_0|U|\psi_0\rangle|$, $|(\langle \psi_0|+\langle\text{Ref}|)U(|\psi_0\rangle+|\text{Ref}\rangle)|/2$, and $|(\langle \psi_0|+\langle\text{Ref}|)U(|\psi_0\rangle+i|\text{Ref}\rangle)|/2$; respectively, these give access to $r$, $\cos(\theta)$, and $\sin(\theta)$.  All three can be measured without adding additional controlled gates and thus without significantly increasing circuit depth.  

One example of a Hamiltonian where this reference state approach is useful is the spin Heisenberg model on a latttice, used for simulating magnetism.  Typical initial states $|\psi_0\rangle$ of interest have total spin 0 (Marshall's theorem, \cite{Auerbach1994}\S5.1), meaning that they are linear combinations of basis states that have an equal number of $|0\rangle$s and $|1\rangle$s.  On the other hand, the fully polarized states $|0\rangle^{\otimes n}$ and $|1\rangle^{\otimes n}$ are exact eigenstates of $H$ with total spin $n/2$ and make for good reference states.  Thus when, e.g., running VQPE to find the ground state of a spin system, we might want to prepare the superposition of the spin-0 initial state $((|01\rangle -|10\rangle)/\sqrt{2})^{\otimes n/2}$ with the reference state $|0\rangle^{\otimes n}$, which is precisely of the form \eqref{eq:two-qubit_chunk_example}.

\subsection{Hamiltonian evolution for certain symmetry sectors}

Related to the previous problem, if we want to apply the operator $e^{-iHt}$ to a given state $|\psi_0\rangle$, we may be able to use multi-state preparation to construct a different operator from $e^{-iHt}$ with a shorter circuit but which still acts correctly on $|\psi_0\rangle$ in particular.

For example, consider a two-qubit system with $H$ given by 
\begin{equation}
    H = \left(\begin{array}{cccc}
        1 &&&\\
        &-1&2&\\
        &2&-1&\\
        &&&1
    \end{array}\right).
\end{equation}
(This is, in fact, the aforementioned spin Heisenberg model, specialized to two spins.)  The time evolution operator $e^{-iHt}$ maintains the same block structure:
\begin{equation}
    e^{-iHt} = \left(\begin{array}{cccc}
        e^{-it} &&&\\
        &e^{it}\cos(2t)&-ie^{it}\sin(2t)&\\
        &-ie^{it}\sin(2t)&e^{it}\cos(2t)&\\
        &&&e^{-it}
    \end{array}\right).
\end{equation}
If the initial state $|\psi_0\rangle$ is of the form $a|01\rangle+b|10\rangle$, the block structure of $e^{-iHt}$ ensures that the time-evolved state will remain a superposition of $|01\rangle$ and $|10\rangle$.  Consequently, we can replace $e^{-iHt}$ by any unitary of the form 
\begin{equation}
    U = \left(\begin{array}{cccc}
        * &&& *\\
        &e^{it}\cos(2t)&-ie^{it}\sin(2t)&\\
        &-ie^{it}\sin(2t)&e^{it}\cos(2t)&\\
        *&&&*
    \end{array}\right)
\end{equation}
without changing the action on $|\psi_0\rangle$.  Thus instead of finding a circuit for $e^{-iHt}$ using unitary synthesis, we can solve the multi-state preparation problem
\begin{subequations}
\begin{align}
    |01\rangle &\mapsto e^{it}\cos(2t)|01\rangle - ie^{it}\sin(2t)|10\rangle\\
    |10\rangle &\mapsto e^{it}\cos(2t)|10\rangle - ie^{it}\sin(2t)|01\rangle,
\end{align}
\end{subequations}
which in general will produce a shorter circuit.

\subsection{Preparation of isometry circuits and quantum channels}

General operations on quantum states are described by quantum channels, i.e. completely positive trace-preserving (CPTP) maps~\cite{nielsen_chuang_2010}, so we would like to be able to find a short and efficient circuit that carries out a given channel.  Although channels are non-unitary, generically involving measurement, reset, and feed-forward, Stinespring's dilation theorem~\cite{Stinespring1955} guarantees that any channel can be implemented by an isometry followed by partial trace, i.e. measurement.  More precisely, a channel from $n_\text{in}$ to $n_\text{out}$ qubits can be implemented by an isometry from $n_\text{in}$ qubits to $n_\text{in} + 2n_\text{out}$ qubits, followed by discarding all but $n_\text{out}$ of the output qubits~\cite{Iten}.  This is the approach to channel compilation taken in, for example, the Mathematica library UniversalQCompiler\cite{ItenCode}.

As we demonstrate in Sec.~\ref{sec:examples} below, our numerical solution to the multi-state preparation problem finds shorter circuits than do existing software implementations for compilation of isometries~\cite{ItenCode, Qiskit, Cirq}.  We will therefore be able to reduce gate depths and thus improve fidelity for simulations of quantum channels.  One key application where this compression will be essential is simulation of open quantum systems, where recent papers have already made use of the representation of a channel as an isometry followed by measurement~\cite{DelRe2020, Hu2020}.

\vspace{0.5cm}

\section{Software implementation and numerical demonstration}\label{sec:examples}

\subsection{BQSKit code}\label{sec:bqskit}

Multi-state preparation is included in the Berkeley Quantum Synthesis Toolkit (BQSKit), an open-source quantum compiler library written in Python and Rust and maintained by Lawrence Berkeley National Laboratory~\cite{BQSKit,BQSKit_code}, beginning with version 1.1.  The problem is specified as a \verb,StateSystem, object, defined using a dictionary where the input states are the keys and the corresponding output states the values, \verb,{psi_in : psi_out},.  Here we include python code for a small example, the two-qubit problem from Eq.~\ref{eq:2-qubit_ex_1}.
\newpage
\begin{small}
\begin{verbatim}
  from bqskit import compile
  from bqskit.qis import StateVector
  from bqskit.qis import StateSystem

  in1 = StateVector(np.array([1,0,0,0]))
  in2 = StateVector(np.array([0,0,0,1]))

  out1 = StateVector(in1)
  out2 = np.array([0,1,-1,0])/np.sqrt(2)
  out2 = StateVector(out2)

  system = StateSystem({in1:out1, in2:out2})

  c = compile(system, optimization_level=2)

  out1_c = c.get_statevector(in1)
  out2_c = c.get_statevector(in2)

  print(c.gate_counts)
  print(out1_c)
  print(out2_c)
  print(out1_c.get_distance_from(out1))
  print(out2_c.get_distance_from(out2))

  >>> {U3Gate: 1, RYGate: 4, CNOTGate: 2}
  >>> [ 1.00e+00+3.06e-07j  1.63e-05+8.47e-12j  
        5.56e-06+3.77e-12j -1.56e-06-1.27e-07j]
  >>> [-7.61e-06-1.54e-12j  7.07e-01+5.03e-08j 
       -7.07e-01+1.29e-07j  7.86e-06+3.48e-12j]
  >>> 2.999380743773372e-10
  >>> 2.644338081836395e-10
\end{verbatim}
\end{small}
The ``off-the-shelf'' \verb,compile, function allows some error in the output vectors as seen here, but the user can request higher precision.

While one solution to this particular problem, the circuit in Eq.~\eqref{eq:ex1_circuit_CH}, could be found relatively easily with pencil and paper, a solution with CNOT as the only two-qubit gate is much more challenging to find by hand; the circuit in Eq.~\eqref{eq:ex1_circuit_CNOT} is the solution generated by the code just above.  Furthermore, the BQSKit implementation of multi-state preparation provides optimal or near-optimal solutions for three- and four-qubit problems that are infeasible to solve by hand.

\subsection{Numerical simulation examples}

In Sec.~\ref{sec:multi-state_def}A, we considered the number of entangling gates such as CNOT and CH required to solve some simple multi-state preparation problems on two qubits.  These sample problems exemplify the theoretically known number of CNOTs required, namely at most one for single-state preparation~\cite{Znidaric2008}, at most two for a 1 qubit $\rightarrow$ 2 qubit isometry~\cite{Iten}, and at most three for a full unitary operation~\cite{Vidal2004, Vatan2004}.

Here we demonstrate our numerical approach on a more difficult case, multi-state preparation for three-qubit and four-qubit states.  We find how the required number of CNOT gates scales with the number of states to simultaneously prepare, and we also show that when the number of states is a power of 2, so that the problem is equivalent to isometry synthesis, we find substantially lower CNOT counts than does the ``top-down'' matrix decomposition approach of \cite{Iten} (used in UniversalQCompiler~\cite{ItenCode}, Qiskit~\cite{Qiskit}, and Cirq~\cite{Cirq}); for large $m$, the reduction is around 30-40\%.

\begin{figure}
    \centering
    \includegraphics[width=0.48\textwidth]{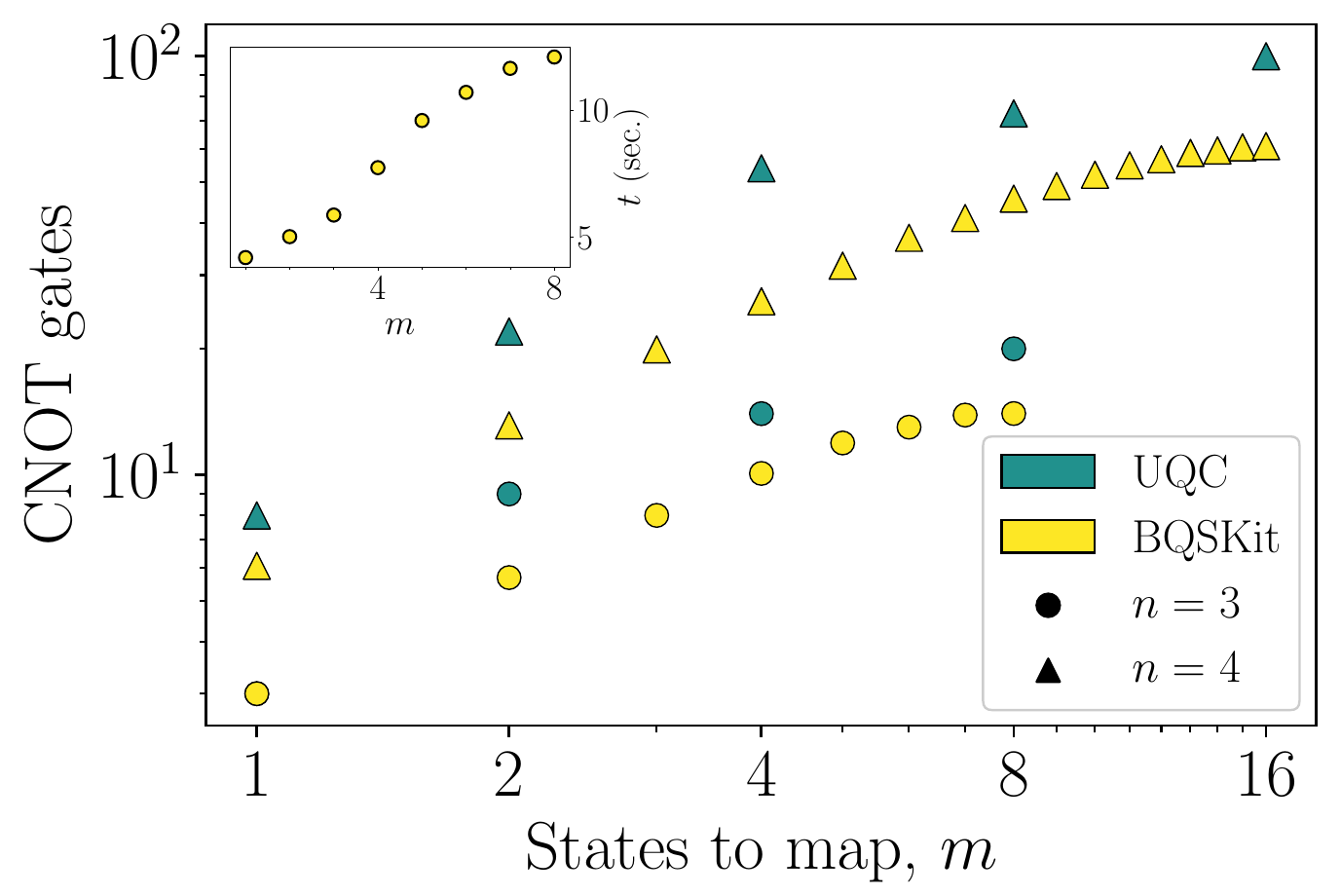}
    \caption{Number of CNOT gates for the shortest circuit found to solve the multi-state mapping problem with 3 (circle markers) and 4 qubits (triangles), and with varying numbers of states to map, $m$.  Yellow points show the CNOT counts for numerical optimization via instantiation with the cost function from Eq.~\eqref{eq:correct_cost_fct} as implemented in BQSKit~\cite{BQSKit_code}, while teal markers show the CNOT counts from the software package UniversalQCompiler~\cite{ItenCode}.  Evidently, the numerical approach finds significantly shorter circuits.  The inset shows the mean computation time to numerically solve the multi-state problem for $n=3$ qubits as a function of the number of states $m$, showing a monotonic and approximately linear dependence.}
    \label{fig:num_expt}
\end{figure}

To be precise, we carry out the following numerical experiment $N_\text{shots}$ times, with $n=3$, 4:
\begin{enumerate}
    \item Generate a $2^n \times 2^n$ unitary operator, $U$.
    \item For $m\in\{1,\cdots,2^n\}$, perform multi-state preparation where the $|v_i\rangle$ are the first $m$ standard basis vectors and $|w_i\rangle$ the first $m$ columns of $U$.  Find the minimum number of CNOT gates needed for each $m$.\footnote{In this numerical experiment, we assume all-to-all connectivity.  BQSKit also supports multi-state preparation with any connectivity graph.}
\end{enumerate}
We show the results of the experiment, both for $n=3$ and $n=4$, in Figure~\ref{fig:num_expt}.  The inset shows how the time required for the compilation scales with the number of states to be mapped, $m$.  Evidently, if one needs a few states to be mapped correctly but does not care about the actions of the circuit on the rest of the Hilbert space, there is substantial freedom to shorten the required circuit by performing mutli-state preparation rather than unitary synthesis, with the side benefit that the classical computation time to find a good circuit will also be reduced.

Recalling Example 1 from Section~\ref{sec:multi-state_def} \textit{A.}, there are two possible reasons for why shorter circuits are returned by our numerical instantiation approach than by top-down matrix decomposition: (a) among all unitary transformations consistent with the multi-state problem, numerical instantiation may be finding one that has a shorter optimal circuit implementation; and/or (b) the top-down approach may be finding a longer than necessary circuit to implement its unitary transformation.  
If only (b) were the case, but not (a), we could feed the output circuits from the top-down approach into a numerical compiler, such as the one included with BQSKit, and find equivalent circuits of the same depth as those from the numerical instantiation approach.  When we perform this additional experiment, we find that, while some circuit depths are indeed reduced, for all cases with $m<2^n$ we still find longer circuits for top-down plus recompiling compared with direct numerical instantiation.  (For example, with $n=4$, $m=2^3$, UniversalQCompiler outputs a circuit with 73 CNOTs, which BQSKit compiles down to 57.  The direct numerical approach gives 46 CNOTs.)  We conclude that the numerical instantiation approach produces fundamentally better solutions that cannot be matched by post-processing the output circuits of the top-down approach.

Finally, we comment on the level of precision in the two approaches.  The matrix decomposition approach returns circuits that produce the desired isometry exactly up to machine precision.  On the other hand, the numerical approach produces circuits that match the desired isometry only up to a specified threshold of the cost function; we currently target a cost threshold that returns isometries accurate to about $10^{-4}$ in each matrix element.  This reduction in precision should not be viewed as a drawback of the method.  Rather, as long as these errors are significantly smaller than those that arise in running the circuit on hardware, the imprecision will have no meaningful impact on the results of real simulations; currently, for a circuit with dozens of two-qubit gates, the hardware errors are orders of magnitude larger.  Furthermore, by allowing some imprecision, it may be possible to produce shorter circuits than if full precision is required, and thus we can choose to slightly increase the error in the theoretical circuit and in exchange substantially lower the noise and infidelity from hardware, a worthwhile tradeoff.


\section{Discussion}\label{sec:discussion}

We have studied a very general quantum circuit synthesis problem that falls between the commonly considered cases of single-state preparation and full unitary synthesis, and which includes each of those problems as a special case.  We prove that a set of input states $\{|v_i\rangle\}$ can be exactly mapped to a set of output states $\{|w_i\rangle\}$ if and only if the overlaps for the two sets of states are equal, $\langle v_i|v_j\rangle = \langle w_i|w_j\rangle$ for all pairs $i$, $j$.  Furthermore, we show that to numerically optimize a quantum circuit to perform the state mapping, a good cost function is given by equation~\eqref{eq:correct_cost_fct}.  We have implemented both the solvability check and the circuit optimization in Berkeley Lab's open-source quantum compiler, BQSKit, and we demonstrated the strong performance of the numerical optimizer on three- and four-qubit problems.

Multi-state preparation is a useful tool for a variety of applications, including preparation of efficient circuits to implement isometries and hence quantum channels.  Other applications include preparation of macroscopic superposition states and reduction of circuits for time evolution in physics and chemistry simulations by preparing a circuit only for the part of the time evolution operator that acts within a specific block of a block-diagonal Hamiltonian.

With the addition of multi-state preparation to BQSKit in version 1.1, we now support the full range of pure-state preparation and unitary synthesis.  Planned future updates to BQSKit will add further capabilities, including direct compilation of non-unitary channels as well as state preparation making use of measurement and reset.  


\section*{Acknowledgment}

We thank Huo Chen for helpful conversations and for pointing out several useful references. 
This work was supported Office of Science, Office of Advanced Scientific Computing Research Accelerated Research for Quantum Computing Program of the U.S. Department of Energy under Contract No. DE-AC02-05CH11231. This research used resources of the National Energy Research Scientific Computing Center (NERSC), a U.S. Department of Energy Office of Science User Facility located at Lawrence Berkeley National Laboratory.


\bibliographystyle{IEEEtran}

\end{document}